# Comparative Stability of Cloned and Non-cloned Code: An Empirical Study

## A Replication Study

Oualid El Halimi, Trith Patel, Zohaib S. Kiyani, Naresh. Kumar, Ankit Singh

Department of Software Engineering, Concordia University, Canada

*Abstract*—Code cloning is an important software engineering aspect. It is a common software reuse principle that consists of duplicating source code within a program or across different systems owned or maintained by the same entity. There are several contradictory claims concerning the impact of cloning on software stability and maintenance effort. Some papers state that cloning is desired since it speeds up the development process and helps stakeholders meet the tight schedule and deliver on time. Other papers argue that code clone leads to code bloat and causes increase software maintenance costs due to copied defects and dead code.

In this paper, we are replicating a previous study done on cloning by Manishankar et al. We are repeating his work using the same methods and metrics but with different subjects and experimenters. The paper we are addressing is "Comparative Stability of Cloned and Non-cloned code: An empirical Study". This latter evaluates the impact of code cloning on code stability using three different stability-measuring methods. Our team will apply the same stability measurement techniques on a different software system developed in C programing language to determine generalizability, assure that the results are reliable, validate their outcomes, and to inspire new search by combining previous findings from related studies.

## 1. INTRODUCTION

Copy-paste is a common activity among software developers. Developers tend to copy a code fragment from one location and paste it into another with or without modifications. This activity leads to the co-existence of multiple code clones in the software systems.

Recently, duplicate code has received a great deal of attention. The common belief is that cloned code makes consistency maintenance task difficult, which introduces bugs and increases maintenance effort. Indeed, to assess the impact of clones on software maintenance phase, the code stability should be investigated. In practice, we need to compare between the code that undergoes cloning and the code that does not. If the cloned code changes more frequently as compared to non-cloned code during the software evolution, we can deduct that cloned code does not really affect maintenance effort. Another stability measurement technique will be used in this paper is the age of change. Meaning, team members will investigate to which extent the age of the system's code fragment influence code stability.

The original research paper written by Manishankar et al. presented an in depth investigation on the comparative stabilities of cloned and non-cloned code. Manishankar's paper illustrated that cloned code exhibits higher changeability than that of non-cloned code. The main objective of our research paper is to repeat Manishankar's et al's study using the same methods and a different software system in an attempt to determine generalizability and ensure that the results are valid and reliable. Our team shall examine a common software system written in C (cUnit: a Unit Testing Framework for C programming language). This latter system has 13 years continuous evolution and 347 active contributors [12].

**Motivation**: Several research efforts have been conducted on the basis that the presence of duplicate code has an influence on the software evolution in modern world. Some papers relate code cloning with system stability and confirm that cloning is beneficial to software development process since it speeds up delivery. Other papers argue that cloning negative effects outweigh the positive ones and lead to extra maintenance efforts and produce system faults and vulnerabilities.

The motivation for this work stems from the need to assess the impact of clones on software maintainability in terms of stability during software evolution. Basically, our team has replicated the work done by Manisankar et al to either refute or validate their findings and demonstrate the effect of cloning on a different software project. Based on the obtained results, we could either confirm the generalizability of Manisankar's work or refute it. Our team will deploy most of the software re-engineering principles taught in the SOEN691-E course and will mostly focus on the "Metrics/reusability of software systems" aspects.

The research questions we address in this paper:

**RQ1: Which type of clone is a significant threat to code stability?**
*Various types of cloning affect system code stability to varying degrees. Our work aims to find out which types of clones have a serious influence on code stability and those with the least effect.*

**RQ2: Do different programming languages exhibit different stability scenarios?**
*Programming languages from distinctive programming paradigms differ in their structures, which can be a cardinal factor in the stability scenarios. Our team will compare our system stability settings to those in the original paper in order*

*to validate Manishankar et al's work. However, due to limited resources and tight schedule, we examine only a C system as a first step to answer this research question.*

**RQ3: Do system size and age affect stability of cloned and non-cloned code?**
*System size and code age should be studied to reveal their contribution to code maintainability. Investigating whether system size and age have a significant impact on the stability of cloned and non-cloned code is one of our primary concerns.*

**Takeaway**: Software Developers can improve the quality of their code by knowing which type of clone could increase the maintenance cost, introduce new bugs and lead to a sloppy design. Thus, knowing which type of clone is harmful would make the system less frequent to changes and add a continuous stability feature.

The remainder of this paper is organized as follows: Section 2 surveys the related work. Section 3 discusses in detail the data that we analyze for our study. Section 4 presents the methods we will adopt. Section 5 conducts the experimental results of our study. Section 6 analyzes the experimental results. Section 7 compares between the three metrics used and explains differences in their results. In section 8, we answer our three research questions, discuss both the findings and their implications for the developers, and we see if our findings generalize. Section 9 concludes the paper.

## 2. BACKGROUND & RELATED WORK

### 2.1 Work related to Code Cloning

There are several debates on the impact of cloning in recent years, which have resulted in a significant study and empirical evidence. Kim et al. [1] introduced clone genealogy model to study clone evolution and the impact of cloning on software systems. To achieve this conclusion, they analyzed two medium-sized Java systems. They showed that (i) aggressive, immediate refactoring may not always improve software quality and (ii) refactoring techniques cannot assist in removing many long-lived, consistently changing clones. They also concluded that there are classes of clones that require different types of maintenance support than conventional refactoring. Saha et al. [2] extended their work by extracting and evaluating code clone genealogies using diverse categories of 17 open source systems at the release level and reported that most of the clones do not require any refactoring effort in the maintenance phase. They have also concluded that it is relatively easy to manage clone in smaller systems than larger ones.

### 2.2 Work related to impact of clones on Code Stability

Impact of clones on code stability is a debating topic and there is research supporting both the views that clones are inherently bad or are conducive to code stability. Lozano and Wermelinger conducted two studies [3, 4] and experimented to assess the effects of clones on the changeability of software using CCFinder [5] as the clone detector. They have calculated three changeability measures: (i) likelihood; (ii) impact of a method change; and, (iii) work required to maintain a method. They conducted two studies and analyzed only a couple of Java systems (4 in [3] and 5 in [4]). They have reported that clones have harmful impact on the maintenance phase because clones often increase maintenance efforts and are vulnerable to system's stability. Juergens et al. [6] studied the impact of clones on large scale commercial systems and suggested that - (i) inconsistent changes occurs frequently with cloned code and (ii) nearly every second unintentional inconsistent change to a clone leads to a fault. Eventually they have concluded that cloning can be a substantial problem during development and maintenance. However Kapser and Godfrey [7] strongly argued against the conventional belief of harmfulness of clones by investigating different cloning patterns. They showed that - (i) about 71% of the cloned code has a kind of positive impact in software maintenance and (ii) cloning can be an effective way of reusing stable and mature features in software evolution. Krinke carried out two case studies. In his first study [8], he did comparative study on cloned and non-cloned code with respect to stability of code. In most recent study [9], he determined the average last change dates of the both cloned and non-cloned code. Both of these studies suggest cloned code to be more stable than non-cloned code. Hotta et al. in a recent study [10], calculated the modification frequencies of cloned and non-cloned code and found that the modification frequency of non-cloned code is higher than that of cloned code. In our study, we have replicated Krinke et al's recent study and used our subject system to reveal more information about comparative stabilities and harmfulness of three clone types along with language based stability trends.

> *Previous research has confirmed the effect of cloned and non-cloned code on the code stability over the years in terms of system size, system age, used programming languages, and clone types. In this paper, we seek to validate previous papers by examining a new system and presenting our findings, which would either validate or refute previous studies. This validation process would help developers to understand the effect of different clone types on system stability and push programmers to take proactive actions to limit system maintenance costs throughout its evolution.*

## 3. STUDY DESIGN

In this section we discuss the subject system used in our study and the collection process for our data.

### 3.1 Data Selection

The system our team has chosen for this study is a lightweight unit-testing framework for C: cUnit. This latter used to write, administer, and run unit tests in systems written in C programming language [12]. It provides C developers and testers with a variety of user interfaces and basic testing functionalities.

One reason to choose this system for our study is the fact that it is commonly used by C programmers and has over 13 years of continuous evolution, which would clearly depict the stability of the system along with its entire life time [13]. Furthermore, cUnit is an open source with more than 347 contributors. Meaning, we have access to the system's version control tool and we could easily examine the code, issue command lines, and compare between files, commits, and revisions of interest.

Another reason to choose cUnit for our subject is the moderate number of revisions (170 revisions); which is feasible for our team to examine during the remaining weeks of the course and, therefore, ensure to meet the delivery deadline.

A brief summary of our subject system set is illustrated in Table 2 below:

Table 2 – Subject Systems

| Language | Systems | Domains | LOC | Revisions |
|---|---|---|---|---|
| **Java** | DNSJava | DNS Protcol | 12621 | 1635 |
| | Ant-Contrib | Web Server | 25092 | 176 |
| | Carol | Game | 79853 | 1699 |
| | Plandora | Management | | 32 |
| **C** | Ctags | Code Generator | 33270 | 774 |
| | TidyFornet | Wrapper | 123409 | 55 |
| | QMailAdmin | Mail | 4054 | 317 |
| | Hash Kill | Password | 83269 | 110 |
| | *cUnit* | *Unit Test* | *17014* | *170* |
| **C#** | GreenShot | Multimedia | 37628 | 999 |
| | ImgSeqScan | Multimedia | 12393 | 73 |
| | CapitalResource | DBMS | 754341 | 122 |
| | MonoOSC | Protocols | 8991 | 355 |

## 3.2 Data Collection

To run our experiments, we need first to detect clones in our Unit Test Framework software. Indeed, we need to detect both exact and near-miss clones at a block level of granularity and at a given function and neglect comments and formatting differences. For this purpose, a clone detection tool developed by Queen's University called *NiCad* is used to detect clones and analyze their stability using a couple of stability dimensions. This tool is designed to implement the NiCad (Automated Detection of Near-Miss International Clones) hybrid [11] clone detection method.

NiCad comes with an easy-to-use command line tool that can be easily embedded in IDE's and other environments. It takes a source directory input and a configuration file specifying the normalizations and filters to be done, and outputs the cloning detection results in either XML or HTML files.

To achieve concrete results, three types of clones should be identified: Type 1, 2, and 3. We set dissimilarity threshold to 0% to types 1 and 2 and 20% to Type 3. Then we set the blind renaming of identifiers to 20% in Type 3.

Dissimilarity Threshold means that in a particular class, clone fragments could have dissimilarities up to a particular threshold value.

Table 1 - NiCad Settings

| Clone Type | Identifier Renaming | Dissimilarity threshold |
|---|---|---|
| Type 1 | none | 0% |
| Type 2 | blindrenaming | 0% |
| Type 3 | blindrenaming | 20% |

## 4. STABILITY MEASURING METHODS

In order to investigate stability feature of our system under study, three methods and associated metrics have been implemented. This section describes both the approaches we will adopt and the metrics to calculate in order to decide on whether cloned code of a subject system is more stable than its non-cloned code. To come up with valuable conclusions, we will compare our stability decisions with other systems and explain the causes of decision dissimilarities.

### 4.1 Modification Frequencies:

The techniques we use to calculate modification frequencies were first defined by Hotta et al[10] study. They have calculated two metrics: (i) $MF_d$ (Modification Frequencies of Duplicate code) and (ii) $MF_n$ (Modification Frequencies of Non-Duplicate code). As a matter of fact, these metrics will be used to estimate the modification frequency of both duplicate and non-duplicate code of our C unit test framework system.

In order to calculate these metrics, we need to exploit source code repository (SourceForge), extract all the revisions from codebase, and examine the differences between consecutive relevant revisions. $MF_d$ and $MF_n$ can be calculated as follows:

$$MF_d = \frac{\sum_{r \epsilon R} MC_d(r)}{|R|} * \frac{\sum_{r \epsilon R} LOC(r)}{\sum_{r \epsilon R} LOC_d(r)} \quad (1)$$

$$MF_n = \frac{\sum_{r \epsilon R} MC_n(r)}{|R|} * \frac{\sum_{r \epsilon R} LOC(r)}{\sum_{r \epsilon R} LOC_n(r)} \quad (2)$$

$MC_d$ and $MC_n$ represent the number of in the duplicate and non-duplicate code regions respectively between revision r and (r+1). R represents the number of revisions of the candidate subject system. LOC(r) represents the number of lines in a given revision r. $LOC_d(r)$ and $LOC_n(r)$ represents correspondingly the number of duplicate and non-duplicate lines of code in a revision r.

### 4.2 Average Last Change Date

In this study, we adopt a second code stability measurement by calculating the last change dates of both cloned and non-cloned code regions. In fact, this concept was introduced by Krinke[9][10] and aims to exploit the SVN directory and use the *blame* command to retrieve each file's revision and the date when each line was last changed. Unlike Hotta's technique that covers all system revisions, Krinke considers only one system revision: the last revision. Indeed, Krinke calculates the average last change dates of cloned and non-cloned code from two level granularities: System level and File level:

**File Level Metrics**: At file granularity level, we need to calculate ($PF_c$) the percentage of files where the cloned code's last change date is older than that of non-cloned code. Furthermore, a second calculation is performed to find ($PF_n$) the percentage of files where the average last change date of cloned code is newer than that of non-cloned code. For both calculations, we examine only the last revision of the Unit Test Framework system. To obtain more accurate results, we considered in our calculations only the analyzable source files. Also, we excluded files with non-cloned code and fully cloned ones.

**System Level Metrics**: At system granularity level, we calculate the average last changed date of cloned code ($ALC_c$) as well as the average last change date of non-cloned code region ($ALC_n$). Again, only the last revision of the subject system is examined.

**Calculation of average last change date**: In order to calculate the average last change, we need to determine the average distance in days from the oldest date. For instance, suppose we have five lines in a file that corresponds to 5 revisions with different revision dates: 01-Jan-11, 05-Jan-11, 08-Jan-11, 12-Jan-11. The average distance in this case would be: (4+7+11+19) / 4 = 10.25. Therefore, starting from 01-Jan-11, the average date is 10.25 days later: 11-Jan-11.

### 4.3 Proposed Variant of Krinke's method

As a matter of fact, the proposed Variant Krinke's method is introduced to cover some of the limitations of the previous Krinke measurement technique such as:

(1) SVN *blame* command gives the revisions and the dates of all the revisions including the blank lines and comments. In fact, Krinke does not exclude comments and blank lines, which may provide inconsistent stability scenarios.

(2) Krinke's method could introduce some rounding errors in its results while calculating the average last change date. In some of the cases, we could end up having equal values of last change dates of cloned and non-cloned code.

While calculating stability scenarios, the proposed Variant Krinke's method overcomes these limitations and provides more accurate data. Indeed, this method does not calculate any file level metrics since the system level files are more suitable in the decision-making. It is important to mention that Hotta's et al method also excludes the blank lines and comments through some preprocessing prior to clone detection.

By calculating their average age, the proposed Variant of Krinke's method is used to analyze the stability of cloned and non-cloned code. In practice, in order to retrieve the age of each cloned and non-cloned lines in our system, the *blame* command line was used.

In this technique, we work on the last revision R of the subject system. We proceed by applying a clone detector on revision R in order to separate the lines of each source file into two sets: (i) a set containing all cloned lines (ii) a set containing all non-cloned lines. In order to calculate the age of a given line of code, we use the *blame* command to assign a revision r to every line x. During the analysis phase, this strategy would help our team to understand that a line x was produced in a revision r and has not been changed up to the last revision R. To calculate this metric, we adopt a variant of Krinke's methodology [10]:

$$age(x) = date(R) - date(revision(x)) \qquad (3)$$

x and r represent the line of code and revision respectively. Krinke has denoted the revision of x as r = revision (x). date(r) represents the creation date of r and date(R) represents the last revision date.

Using this formula, we can find the age (in days) of a given line of code by subtracting the revision date of a given line of code from the last revision date. Therefore, we will be able to calculate the following two averages: ($AA_c$) Average Age of cloned code from the subject system and ($AA_n$) Average Age of non-cloned. It is important to mention that we cover all lines of all source files of our system to calculate these two averages.

### 4.4 Major Differences Between Hotta's Method and The Other Two Methods

The three adopted methods in this study aim at evaluating the code stability of the subject system. However, there are major differences between them. One of the differences between Hotta et al's method and the other two is that Hotta's method considers all changes to a code region from its creation regardless of when these changes are applied. Meanwhile, the other two methods consider only the last modification and neglect previous code changes in earlier revisions. As an example, suppose a file has two lines x and y at revision 1. Assume this file went through 100 commits, during which, line x knew 5 changes and y only one change. Let the change on x took place at the 50th commit and the last change of y at the 99[th] commit. Issuing a *blame* command on the last revision (100) of the file will assign x with revision 50 and y with 99. In this case, according to both Krinke's method and the variant, the line x is older than the line y because the revision date of line x is older than the revision date corresponding to line y. Therefore, x is considered to be more stable than y by these two methods. On the other hand, Hotta et al's method counts the number of modifications that took place between these two lines. As a result, Hotta et al assumes that y is more stable than x since the modification frequency of x are bigger than that of y.

### 5. EXPERIMENTAL RESULTS

Our team has examined the cUnit source code and applied the 3 stability measurements by considering the three types of clones (Type-1, Type-2 and Type-3). The following table-3 summarizes the Krinke decision-making strategy to determine code stability criteria for different scenarios.

**Table 3 - Krinke decision-making strategy table**

| Method | Metrics | | Decision Making | |
|---|---|---|---|---|
| | CC | NC | CC More Stable | NC More Stable |
| Hotta et al. [6] | $MF_d$ | $MF_n$ | $MF_d < MF_n$ | $MF_n < MF_d$ |
| Krinke [12] | $ALC_c$, $PF_c$ | $ALC_n$, $PF_n$ | $ALC_c$ is older | $ALC_n$ is older |
| Krinke's Variant | $AA_c$ | $AA_n$ | $AA_c > AA_n$ | $AA_n > AA_c$ |

Special Decision for Krinke's Method
if $ALC_c = ALC_n$ then
$PF_c > PF_n$ implies CC more stable
$PF_n > PF_c$ implies NC more stable
CC = Cloned Code        NC = Non-cloned Code

Table 4 shows the average last change dates obtained by applying Krinke's method. We have calculated the Average Last Change date of both cloned ($ALC_c$) and non-cloned ($ALC_n$) code regions for the three type clones (Types-1, Type-2, and Type-3). Then, the modification frequencies were calculated for our unit test framework system. We have applied the two modification frequencies formulas described in section 3.2 to all cUnit system revisions and for each of the clone types: 1,2, and 3.

Table 5 summarizes the results obtained from calculating the modification frequencies of cloned ($MF_d$) and non-cloned ($MF_n$) code by Hotta's et al method.

**Table 5 - Modification Frequencies of Cloned ($MF_d$) and non-cloned ($MF_n$) code by Hotta's et al**

| Lang. | Systems | Type-1 | | Type-2 | | Type-3 | |
|---|---|---|---|---|---|---|---|
| | | $MF_d$ | $MF_n$ | $MF_d$ | $MF_n$ | $MF_d$ | $MF_n$ |
| Java | DNSJava | 21.61 | 7.12 | 19.34 | 6.99 | 7.93 | 8.66 |
| | Ant-Contrib | 3.62 | 1.49 | 2.02 | 1.52 | 1.43 | 1.59 |
| | Carol | 8.15 | 6.60 | 4.07 | 3.69 | 9.91 | 8.97 |
| | Plandora | 0.44 | 0.92 | 0.45 | 0.97 | 0.55 | 1.11 |
| C | Ctags | 6.37 | 3.82 | 7.19 | 7.17 | 6.71 | 3.68 |
| | TidyFornet | 0 | 5.07 | 0 | 4.87 | 0 | 4.89 |
| | QMailAdmin | 5.09 | 2.74 | 8.83 | 5.47 | 8.24 | 2.58 |
| | Hash Kill | 61.24 | 115.22 | 59.92 | 115.64 | 65.75 | 118 |
| | *cUnit* | *1.92* | *0.32* | *0.76* | *0.40* | *0.62* | *0.46* |
| C# | GreenShot | 7.94 | 6.07 | 6.92 | 6.07 | 8.13 | 6.06 |
| | ImgSeqScan | 0 | 20.93 | 0 | 21.06 | 0 | 21.29 |
| | CapitalResource | 0 | 67.15 | 0 | 67.31 | 3.63 | 67.11 |
| | MonoOSC | 8.58 | 29.14 | 7.92 | 29.23 | 10.62 | 29.63 |

The average days of cloned ($AA_c$) and non-cloned ($AA_n$) code were calculated using the proposed variant for the three clone types: Type-1, Type-2, and Type-3. The method is described in section 3.3. Average Dates obtained for cUnit are illustrated in Table 6.

**Table 6 – Average days in days of Cloned ($AA_c$) and non-cloned ($AA_n$) code by he proposed variant**

| Lang. | Systems | Type-1 | | Type-2 | | Type-3 | |
|---|---|---|---|---|---|---|---|
| | | $AA_c$ | $AA_n$ | $AA_c$ | $AA_n$ | $AA_c$ | $AA_n$ |
| Java | DNSJava | 2181 | 2441 | 2247 | 2243 | 2210 | 2446 |
| | Ant-Contrib | 853.6 | 903.7 | 896.1 | 903.3 | 870.6 | 904.4 |
| | Carol | 189.6 | 210.9 | 190.3 | 211.3 | 227 | 209.6 |
| | Plandora | 51.82 | 51.32 | 50.6 | 51.4 | 51.5 | 51.3 |
| C | Ctags | 1301.4 | 1345 | 1351 | 1345 | 1564 | 1343 |
| | TidyFornet | 104.5 | 97.9 | 84.9 | 98.1 | 72.8 | 98.3 |
| | QMailAdmin | 2664.2 | 2678 | 2651 | 2678 | 2644 | 2678 |
| | Hash Kill | 261.5 | 118.5 | 250.3 | 118.4 | 257.9 | 118 |
| | *cUnit* | *1564* | *1777.4* | *2076* | *1754.3* | *2321.3* | *2287* |
| C# | GreenShot | 103.1 | 97.1 | 102.9 | 97.1 | 94.5 | 97.2 |
| | ImgSeqScan | 14 | 20 | 15.6 | 20.3 | 14.4 | 20.4 |
| | CapitalResource | 86.7 | 86.5 | 88 | 86.5 | 89.3 | 86.5 |
| | MonoOSC | 315.4 | 313.5 | 347.9 | 313 | 378 | 312.3 |

Indeed, Decision making strategies for Tables 4, 5 and 6 are depicted in Table 3.

## 6. EXPERIMENTAL RESULTS ANALYSIS

Throughout this section of the paper, we address the three research questions introduced earlier in the first section of this paper. In fact, having all these results in Tables 4, 5 and 6, we need a strategy to deduct which code is more stable than the other. For this purpose, a stability decision table is used to help compare all the stability scenarios. We used symbols to ease the reading of the table: (+) implies that the cloned code is more stable while (-) implies that the non-cloned is more stable. Table 7 below summarizes the results. It contains the decisions of 117 decision points: 13 subject systems, three measurement methods, and three cloning types for each measurement.

**Table 7 – Comparative Stability Scenarios**

| Lang. | Methods | Krinke [8] | | | Hotta [10] | | | Variant | | |
|---|---|---|---|---|---|---|---|---|---|---|
| | Clone Types | Type-1 | Type-2 | Type-3 | Type-1 | Type-2 | Type-3 | Type-1 | Type-2 | Type-3 |
| Java | DNSJava | (-) | (-) | (-) | (-) | (-) | ⊕ | (-) | (-) | (-) |
| | Ant-Contrib | (-) | (-) | (-) | (-) | (-) | ⊕ | (-) | (-) | (-) |
| | Carol | (-) | (-) | ⊕ | (-) | (-) | (-) | (-) | (-) | ⊕ |
| | Plandora | ⊕ | ⊕ | ⊕ | ⊕ | ⊕ | ⊕ | ⊕ | (-) | ⊕ |
| C | Ctags | (-) | (-) | ⊕ | (-) | (-) | (-) | (-) | ⊕ | ⊕ |
| | TidyFornet | ⊕ | (-) | (-) | ⊕ | ⊕ | ⊕ | ⊕ | (-) | (-) |
| | QMailAdmin | (-) | (-) | (-) | (-) | (-) | (-) | (-) | (-) | (-) |
| | Hash Kill | ⊕ | ⊕ | ⊕ | ⊕ | ⊕ | ⊕ | ⊕ | ⊕ | ⊕ |
| | *cUnit* | (-) | (-) | ⊕ | (-) | (-) | (-) | (-) | ⊕ | ⊕ |
| C# | GreenShot | ⊕ | ⊕ | ⊕ | (-) | (-) | (-) | ⊕ | ⊕ | (-) |
| | ImgSeqScan | (-) | (-) | (-) | ⊕ | ⊕ | ⊕ | (-) | (-) | (-) |
| | CapitalResource | (-) | ⊕ | ⊕ | ⊕ | ⊕ | ⊕ | ⊕ | ⊕ | ⊕ |
| | MonoOSC | (-) | ⊕ | ⊕ | ⊕ | ⊕ | ⊕ | ⊕ | ⊕ | ⊕ |
| ⊕ = Cloned code more stable | | | | | | | | | | |
| (-) = Non-cloned code more stable | | | | | | | | | | |

Table 4 – Average Last Change Dates of Cloned (ALC$_c$) and Non-Cloned(ALC$_n$) code

| Prog. Language | Systems | Type-1 | | Type-2 | | Type-3 | |
|---|---|---|---|---|---|---|---|
| | | ALC$_c$ | ALC$_n$ | ALC$_c$ | ALC$_n$ | ALC$_c$ | ALC$_n$ |
| Java | DNSJava | 24-Mar-05 | 26-Apr-04 | 21-Jan-05 | 24-Apr-04 | 11-Jun-10 | 21-Jun-10 |
| | Ant-Contrib | 26-Apr-04 | 03-Aug-06 | 18-Sep-06 | 02-Aug-06 | 19-Jan-11 | 14-Jan-11 |
| | Carol | 21-Jan-05 | 18-Jan-07 | 25-Nov-07 | 14-Jan-07 | 13-Dec-08 | 12-Dec-08 |
| | Plandora | 31-Jan-11 | 01-Feb-11 | 01-Feb-11 | 01-Feb-11 | 08-Apr-09 | 21-Mar-09 |
| C | Ctags | 27-May-07 | 31-Dec-06 | 24-Mar-07 | 31-Dec-07 | 12-Jun-10 | 21-Mar-09 |
| | TidyFornet | 10-Jan-07 | 16-Jan-07 | 29-Jan-07 | 16-Jan-07 | 17-Jan-11 | 14-Jan-11 |
| | QMailAdmin | 07-Nov-03 | 24-Oct-03 | 19-Nov-03 | 24-Oct-03 | 11-Dec-08 | 12-Dec-08 |
| | Hash Kill | 14-Jul-10 | 02-Dec-10 | 27-Jul-10 | 02-Dec-10 | 05-Mar-09 | 21-Mar-09 |
| | *cUnit* | *11-Jun-09* | *16-Dec-08* | *18-Jan-09* | *25-Nov-08* | *06-Jan-10* | *19-Feb-10* |
| C# | GreenShot | 11-Jun-10 | 21-Jun-10 | 12-Jun-10 | 21-Jun-10 | 20-Jun-10 | 20-Jun-10 |
| | ImgSeqScan | 19-Jan-11 | 21-Jun-10 | 17-Jan-11 | 17-Jan-11 | 19-Jan-11 | 14-Jan-11 |
| | CapitalResource | 13-Dec-08 | 14-Jan-11 | 11-Dec-08 | 11-Dec-08 | 10-Dec-08 | 12-Dec-08 |
| | MonoOSC | 08-Apr-09 | 21-Mar-09 | 05-Mar-09 | 05-Mar-09 | 21-Jan-09 | 22-Mar-09 |

Examining table 7 leads us to build Table 8, which is used to summarize the overall stability decisions of all the methods. This table contains 39 decision points (13 subject systems and 3 clone types) and each table cell represents a decision and implies the agreements (-) or (+) as well as the disagreement (X) of the candidate methods. Indeed, the decisions of candidate methods of Type-1 clones of cUnit are similar (+). For the Type-2 case, the proposed variant agrees with Hotta's method and disagrees with Krinke's. Finally, for Type-3, the proposed variant agrees with Krinke's method and disagrees with Hotta's. Table 8 below summarizes subject systems' stability decisions of all the methods.

Table 8 – Overall stability decisions by methods

| Systems | Java | | | | C | | | | | C# | | | |
|---|---|---|---|---|---|---|---|---|---|---|---|---|---|
| Clones Types | DNSJava | Ant-Contrib | Carol | Plandora | Ctags | TidyFornet | QMailAdmin | Hash Kill | *cUnit* | GreenShot | ImgSeqScan | CapitalResource | MonoOSC |
| Type-1 | - | - | - | ⊕ | - | ⊕ | - | ⊕ | - | ⊗ | ⊗ | ⊗ | ⊗ |
| Type-2 | - | - | - | ⊗ | ⊗ | ⊗ | - | ⊕ | ⊗ | ⊗ | ⊗ | ⊕ | ⊕ |
| Type-3 | ⊗ | ⊗ | ⊗ | ⊕ | ⊗ | ⊗ | - | ⊕ | ⊗ | ⊗ | ⊗ | ⊕ | ⊕ |

⊕ = All methods agree that cloned code is more stable
(-) = All methods agree that non-cloned code is more stable
⊗ = Not all of the methods agreed

We can deduct from this table 8 that non-cloned code is more stable for the cUnit system for Type-1. Nonetheless, we cannot generalize our results to deduct stability for Type-2 and Type-3 since not all of the 3 methods agree.

## 7. METHODS ANALYSIS

In this section we will provide the analysis of systems code stability with regard to the methods used. Table 6 represents the results obtained by examining 13 systems and 3 clone types, which result to 117 decision points where each of the three methods contributes with 36 decisions. Table 9 below processes the decision-making scenarios presented in Table 7:

Table 9- Stability of candidate methods

| Decision Parameters | % of Decision Points | | |
|---|---|---|---|
| | Krinke[8] | Hotta [10] | Variant |
| Cloned code more stable | 43.58 | 51.28 | 48.71 |
| Non-Cloned code more stable | 56.41 | 48.71 | 51.28 |

Table 9 contains quite similar statistics for Krinke's method and the variant. However, Hotta's method produces larger variations from the other two methods. In general, Krinke's and the variant approaches suggest that non-cloned code is more stable while Hotta's approach suggests that clones code is the one the most stable.

By examining the decision stabilities of the three methods (Table 8), 18 cells are in disagreements among the 39 candidate methods. This mean, methods disagree for 46.15% of the cases. Our subject system 'cUnit' has 67% disagreement since all the three methods do not agree about clone stability of both Type-2 and Type-3. On the other hand, there are two systems (Hash Kill and QMail) where all the methods agree on their decisions. Nevertheless, we have other systems with strong disagreements like 'GreenShot' and 'ImgSeqScan' systems due to the major differences in decision strategies explained earlier in section 4.4.

**Analysis of Strong Disagreements**: Some of the candidate methods show a strong disagreement when applied on the systems under study. Let us take 'ImgSeqScan as an example. For the 3 clone types, there is a strong disagreement to the three methods decisions. Hotta et al method shows that the three clone types indicate that the clone code regions are more stable than non-cloned ones (Table 7). Meanwhile, Both Krinke and Variant methods indicate contradictory decisions. More importantly, by examining Table 5, 'ImgSeqScan' has a 0 $MF_d$ value for all the three clone types. This means that the duplicate source code did not undergo any modification during its entire life-time (73 total commits). If the blame command is issued, it will retrieve the creation dates of the cloned code fragments in the last system revision. Based on these dates, Krinke's method could calculate the average date for the cloned regions. If the creation date of the clone code sections are newer than the creation dates of non-clone code, the average last change date of the clone region would be newer than non-clone's. As a result, Krinke's method considers non-clone code to be more stable than the cloned code. This example explains the strong disagreement between the applied methods since the cloned/non-cloned code fragments of a given system could be considered less-stable than its counterpart even if the system did not experience any modification during its evolution while its counterpart does.

Based on this analysis, we have 46.15% dissimilarities in table 8. Besides, Hotta et al's method has strong disagreement with Krinke's and Variant methods in several cases.

**Results of the Analysis**: By examining the above analysis, we can clearly state that the cloned code has higher probability to put the system to an unstable state compared to the probability of non-cloned code. Among 117 cells in table 7, cloned code is less stable than non-cloned code for 61 cells. Therefore, the probability of cloned code making the system unstable is 52%, which outweighs the probability of non-cloned code (48%).

## 8. ANSWERS TO RESEARCH QUESTIONS

Now that we have analyzed the results obtained from the experimental results, this section answers the research questions by investigating how stability decision of a subject system varies according to the clone type, programming language, system size and age.

**RQ1: Which type of clone is a significant threat to code stability?**

**Motivation:** As mentioned earlier, code cloning is a common aspect in the software engineering process. Therefore, developers need to consider which clone type is more harmful to software systems stability in order to limit cloned bugs and reduce maintenance efforts.

**Approach:** To answer this question, we examine both the comparative study stability scenarios table (Table 7) and the overall stability decisions with regard to methods (Table 9). From these two tables, we extract the stability ratios of different clone types for all the systems under this study and compare this data with the findings of the original study.

**Findings:** By analyzing Table 7, we can deduct that the stability decision made by a method on a specific subject system by taking into consideration the three clone types is similar for 25 cases. This implies a percentage of 64.10% cases among 39 cases. Indeed, each of these 14 (39-25 = 14) cases consists of three decisions made by a particular method on a given system with some minor variations. As an example, let us consider the decisions made by Hotta's method for 'cUnit'. For Type-1, Type2, and Type-3 cases (Table 5), $MF_d > MF_n$ suggests that three type clones are less stable than the non-cloned code. On the other hand, Krinke's method states that Type-1 and Type-2 clones are less stable, unlike Type-3 clone (more stable). Also, it is important to mention that the difference between $ALC_n$ and $ALC_C$ for Type-3 case is smaller compared to the other cases.

In fact, the analysis of each clone type in table 8 leads to find some of the stability statistics, which are illustrated in table 9. This latter indicates the ratio of code stability with regard to clone type.

(i) For the subject systems, the percentage of subject systems where cloned code is more stable than non-cloned code is growing.

(ii) For the subject systems, the percentage of subject systems where cloned code is less stable than non-cloned code is quickly decreasing.

(iii) For each clone type, the proportion of disagreements is increasing.

**Table 10 – Stability of different clone types**

| Decision Parameters | % of Subject System | | |
|---|---|---|---|
| | Type-1 | Type-2 | Type-3 |
| **Cloned code more stable** | 23.07 | 23.07 | 30.76 |
| **Non-Cloned code more stable** | 46.13 | 30.76 | 7.69 |
| **Conflicting** | 30.76 | 46.15 | 61.53 |

The analysis shows that exact clones (Type-1 clones) and the clones with differences in data types and identifier names (Type-2 clones) represent the clones the most harmful for a system because they cause the system to be more unstable than the non-cloned code.

By examining the agreed decision points in table 8, we can notice that in Type-1 case:

(i) Code clones make system less stable with a probability of Number of cells with cloned code less stable/total number of cells = 6/13= 0.46

(ii) Non-cloned code makes the system less stable with a probability = 3/13 = 0.23

The examination of Type-2 clones case leads to two probabilities: 4/13 = 0.30 percent probability that code clones make system less stable and 3/13= 0.23 percent probability that non-cloned code makes the system less stable.
The examination of type-3 clones case leads to opposite results. Type 3 clones decrease stability with 1/13= 0.076 percent, which is less than the probability of non-cloned code (0.30). Therefore, Type-1 and Type-2 cloned code cases have higher probability of decreasing the stability of the system while Type-3 case has less probability to decrease systems stability.

Finally, by comparing the stability of different clone types of this study (table 9) with the original study (Manishankar el al's table 11), our study reveals that the percentage of 'non-cloned code more stable' for Type-1 (46.13%) is bigger than the finding of the original paper for the same clone type (41.67%). However, for Type-2, our study shows that the percentage of 'non-cloned code more stable' (30.76%) is less than Manishankar et al's finding (33.33%).

As a result, our study validates Manishankar's et al's work and strengthens the theory that Type-1 non-cloned code is more stable than cloned code over the evolution of software systems. However, our study does not fully validate the theory that Type-2 clone code makes the system more stable.

**Discussion:** Types-1 cloned code fragments threaten the stability of software systems. So, developers should put more attention to exact copy-paste activities during the development process. Type-2 clones could be threatening too but not as much as Type-1 clones. So, the process of changing data types and renaming identifiers during the development phase is not enough to ensure better code stability over the years.

The different finding our team has obtained concerning Type-2 cloned could be related to several factors. One of which is cUnit reviewers committee has a strict policy to limit contributors from copying and modifying the code. Another factor could be that cUnit contributors are creative and they write the code themselves rather than lazily copy and then modify cUnit code fragments.

**RQ2: Do different programming languages exhibit different stability scenarios?**

In this section, we will analyze our obtained results to see if there is a correlation between code stability and the system's programming language.

**Motivation:** The main motivation of answering this research question is to find if there is a connection between system's code stability and used programming languages. Said otherwise: To which extent a given programming language or a programming paradigm could influence system stability over its lifetime. The results would definitely help software developers decide what language to use to build their systems especially if they anticipate several modules/components with repeated code sections.

**Approach:** To examine the influence of programming languages on the code stability, we need to extract from Table 8 the percentage of agreed decision points of stable cloned codes, stable non-cloned code, as well as the conflicting decisions. Then, we compare these percentages with the Manishankar et al's finding. During this study, we could examine only one system written in C (cUnit) due to time constraints and limited resources. Therefore, we could only validate the stability scenario of C programming language.

Furthermore, it is important to mention that Manishankar found that the percentage of decision points of C language yield to similar results for all the 3 decision parameters (33.33% for each decision parameter), which will be easy to validate by introducing our cUnit system to the subject systems data set.

**Findings**: Our set of data consists of 13 systems developed in three different programming languages. According to table 10, each language contributes to a couple of decision points, as illustrated in Table 11 below:

Table 11- Stability of programming languages

| Decision Parameters | % of agreed Decision Points | | |
|---|---|---|---|
| | Java | C | C# |
| **Cloned code more stable** | 16.66 | 26.66 | 33.33 |
| **Non-Cloned code more stable** | 50 | 33.33 | 0 |
| **Conflicting** | 33.33 | 40 | 66.67 |

By examining table 11, the current statistics show that Java systems have the least proportion of decision disagreement followed by C systems while C# systems exhibit the highest rate.

Compared to Mankanshar's C language findings, our study reveals similar percentage of agreed decision points value of having "non-cloned code more stable" (33.33%). However, the probability of conflicting decision outweighs Mankanshar's since the "Cloned code more stable" decision parameter has decreased by 6.67%, which explains the increase of the conflicting decision parameter by 6.67%

**Discussion:** At this point, we can conclude that clones in C systems exhibit high instabilities. Therefore, C developers should be more careful when they perform code cloning during the development process. Nonetheless, given the absence of Java and C# systems in our study, we are not able to validate neither refute Mankanshar's theory concerning the contribution of Java and C# to the stability scenarios. Our team will investigate on this case in a future work.

**RQ3: Do system size and age affect stability of cloned and non-cloned code?**

In this section, we will analyze whether system age and size have an impact on the comparative stabilities. This is done by monitoring how modifications occur in the cloned and non-cloned code segments as the system ages.

**Motivation:** Developers and system maintainers should know if there is a correlation between system age, system size, and code instability. This way, they could be aware of how to deal with old and/or complex systems consisting of cloned code portions. Also, software architects should be mindful of the impact of code cloning if there is a need to rejuvenate legacy systems framework through reengineering, for instance.

**Approach:** We select 'cUnit', 'DNSJava', and 'Carol', as system candidates for this investigation. These three selected systems belong to different system sizes/ages and yield contradictory stability scenarios for Hotta's method. Based on the cUnit calculated modification frequencies, we plot a graph representing these MFs with regard to the number of revisions.

**Findings:** Compared to 'cUnit' subject system, 'Carol' and 'DNSJava' have larger number of revisions. In terms of lines of code (LOC), 'cUnit' has the smallest number (17 014 LOC) followed by DNSJava while 'CAROL' has the largest LOC value (23 373 LOC).

The following figures below (Figure 1,2, and 3) depict the modification frequencies of each revision. Figures 1 and 2 were taken from Manishankar et al's paper. Our team has plotted calculated cUnit Type-1 modification frequencies according to Hotta et al's method to produce figure 3. The calculations covered all the cUnit revisions. (1 to 170)

**Figure 1: MFs for Carol (Type-3)**

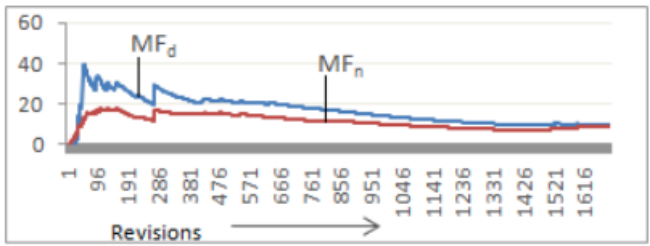

**Figure 2: MFs for DNSJava (Type-1)**

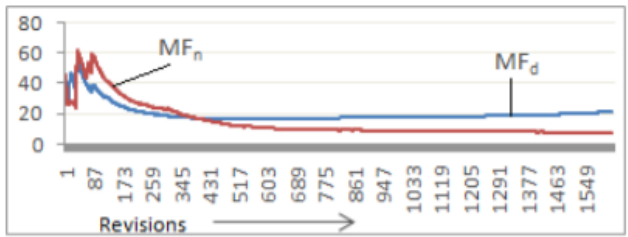

**Figure 3: MFs for cUnit (Type-1)**

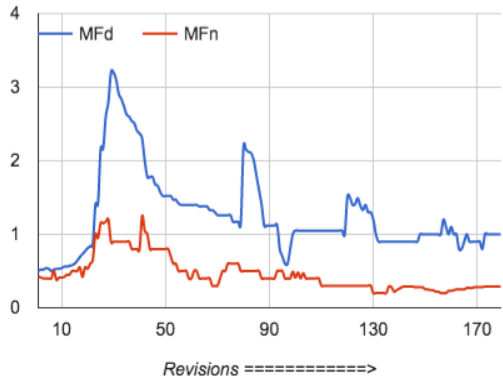

By observing Graphs 1, 2 and 3, we notice there is no consistent change pattern among the MFs of Carol and DNSJava systems. Over the revisions, Modification frequencies of duplicate code are higher in general than the modification frequencies of non-duplicate code. Therefore, we can say that system size and age do not affect the stability of cloned and non-cloned code, which validates Manishankar's study.

**Discussion:** There could be many explanations to the results obtained. One could argue that internal company's development strategy would lead to such high modification frequencies in cloned code. In fact, some programmers might be afraid to modify already reviewed code. So, they tend to create copies in order to apply changes instead of modifying code owned by other authors. Besides, due to the absence of experienced senior software developers, some novice programmers tend to create clone code to speed up development process, which results in bugs propagation, frequent modifications, and ceaseless bug fixes.

## 9. CONCLUSION

The main objective of this study is to help software programmers understand the effect of code cloning in evolving software systems and prioritize their development efforts to reduce maintenance costs.

In this paper, we presented an in-depth investigation on the comparative stabilities of cloned and non-cloned code to validate Manishankar et al' finding by answering three research questions. Indeed, the examination of our C subject system reveals that exact clones (Type-1) are very harmful for the code stability. Yet, clones with different data types and identifier names (Type-2) are less severe than Type-1. Compared to non-cloned code, Types-1 and 2 clones exhibit higher probability values of instabilities, which validate Mankashar's theory about clone type impact on code stability. Thus, these clones should be given more attention from the development and maintenance perspectives.

Our analysis of different systems (cUnit, DNSJava, and Carol) belonging to varying LOC and revision dates indicate that both size and age do not affect code stability since modification frequencies of duplicate code are high over the years, which validates Manishankar's theory concerning the impact of system age and size on stability of cloned code.

On the other hand, the examination cUnit shows that C language has high percentage of agreed decision points and high modification probabilities. So, C systems exhibit high instabilities. At this stage, our team has partially validated Manishankar's theory for C language (not for Java).

Our future plan is to perform empirical studies using a wider range of subject systems belonging to different programing language to fully validate our research question 2.

## 9. ACKNOWLEDGEMNT

We hereby acknowledge our professor Dr. Shihab Emad for his guidance and dedication to make this project a success and to all team members.